# Phase Transition in unrestricted random SAT

Bernd R. Schuh                                            bernd.schuh@netcologne.de

**Abstract.**

For random CNF formulae with m clauses, n variables and an unrestricted number of literals per clause the transition from high to low satisfiability can be determined exactly for large n. The critical density m/n turns out to be strongly n-dependent, $c_{cr}= \ln(2)/(1-p)^n$, where pn is the mean number of positive literals per clause. This is in contrast to restricted random SAT problems (random K-SAT), where the critical ratio m/n is a constant. In a biased model, where variables $a_j$ und their complements $\bar{a}_j$ occur with different probabilities $p > q$, the critical line which separates the (m,n)-plane into regions of high and low satisfiability lies wthin a narrow strip between lower and upper bounds given by $m_{lb}=n\ln(2)(1-q)^{-n}$ and $m_{ub}=n\ln(2)(1-p)^{-n}$. All transition lines are calculated by the second moment method applied to the number of solutions N of a formula. Again in contrast to random K-SAT, the method does not fail for the unrestricted model, and it is not necessary to construct custom made order parameters. It is argued that the difference to K-SAT stems from long range interactions between solutions which are not cut off by disorder. We also point out that models with a fixed number of literals per clause, i.e. random K-SAT, may give restricted information on correlations in solution space, because they suffer from a limited sample space: the set of all K-SAT CNF-formulae with n variables contains only a tiny fraction of all possible logically inequivalent formulae with n variables.

## 1. Introduction.

Complex computational problems can be mapped to corresponding satisfiability problems in propositional logic. Such SAT-problems are conveniently formulated in conjunctional normal form (CNF), i.e. a set of constraints given as a disjunction of basic variables and their negation. A couple of powerful algorithms exist for solving CNF-SAT problems. Most investigations concentrate on K-SAT where each clause contains exactly K literals. Since K-SAT is NP-complete für all K≥ 3 and probably NP ≠ P (although unproven) one can expect that at least some problems have exponential running times under any algorithm. Much effort has been invested in identifying classes of such hard solvable problems. One approach is the investigation of random K-SAT. It is well known that random K-SAT problems undergo a relatively sharp transition from satisfiability to non-satisfiability beyond a critical value m/n = $c_{cr}(K)$, where m is the number of clauses, n the number of variables. For an overview see [1,2]. The critical region has been investigated intensely to get more insight into the nature of hard problems which one expects to occur close to the transition line, see , e.g., [7,8]. It turns out that solutions, i.e. satisfying assignments get sparse in the sense that they become uncorrelated for larger (hamming) distances. For the same reason it is difficult to derive rigorous results on the transition region, in particular to locate the transition line exactly.

We revisit a model where the number of literals per clause is not restricted to a fixed value. Instead each variable and its negation in a randomly and independently generated clause may occur with probability p, such that 2p is the probable fraction of literals in a random clause. Computer experiments on similar models have identified regions of low and high satisfiability depending on the mean clause length κ = 2pn, [3]. Such (n,m,p)-models have been abandoned earlier on because they did not yield valuable information on the location of hard solvable problems in the space of solutions. Also, models with mixed clause lengths are known to exhibit less hard instances than those with fixed K. On the other hand, they are interesting because the hardest problems of random SAT with mixed clause lengths can be magnitudes harder than those of fixed K models, [5]. Our reason to investigate the model is twofold: first we would like to point out that a relatively sharp transition can be determined rigorously for (n,m,p)-models, which seems to be a novel result. Second we point to the profound difference of these results to random K-SAT and try to explain these differences.

As for the methods to determine the transition line we make use of the same methods that have become fashionable for K-SAT, i.e. generating lower and upper bounds (threshholds) on the critical density via the first and second moment of some "order parameter" like e.g. the number of solutions. This approach has lead to nearly matching threshholds, not yet exact ones, for models with fixed K, [4]. In the case of the unbiased unrestricted model considered here it does lead to matching threshholds and thus to a rigorous calculation of the transition line.

*Preliminary considerations.*

A formula F of propositional calculus with n variables $a_i$ and m clauses $C_j$ in conjunctive normal form (CNF), i.e. a set $\{C_1,...,C_m\}$ of clauses, which consist of n literals $l_s$ each. The literals can be a variable *a*, its complement *ā* or falsum *0* which is unsatisfiable under all truth assignments by definition.

A truth assignment T: $\{a_1,...,a_n\} \longrightarrow \{0,1\}$ satisfies a literal $l_s$ iff $T(l_s) = 1$. By definition $T(0)=0$ and $T(\bar{a}) = 1-T(a)$. A clause is said to be satisfied by a truth assignment T iff T satisfies any of its literals. T is said to satisfy a CNF formula F iff all clauses of F are satisfied. Then T is called a *solution* of F. One writes $T(F) = 1$ iff T is a solution, $T(F)=0$ otherwise.

A formula F is said to be satisfiable or SAT iff it has at least one solution. Otherwise it is unsatisfiable or UNSAT.

For n variables there exist $2^n$ distinct truth assignments T which we label by the $2^n$ n-tupels

(1)    $\boldsymbol{\tau}=(\tau_1,...,\tau_n) \;\varepsilon\; \{1,-1\}^n$,  $\tau_s = 2\,T(a_s) - 1$.

For some calculations it is convenient to represent a CNF-formula F by its clause variable matrix $f_{js}$

(2)    $\begin{aligned} f_{js} &= \;\;\;1 \text{ if } C_j \text{ contains } a_s \\ f_{js} &= -1 \text{ if } C_j \text{ contains } \bar{a}_s \\ f_{js} &= \;\;\;0 \text{ if } C_j \text{ contains neither } a_s \text{ nor } \bar{a}_s \end{aligned}$

Then T(F) can be written as

(3)    $T_{\boldsymbol{\tau}}(F) = \prod_{j=1}^{m} [1 - \prod_{s=1}^{n} (1-\delta(\tau_s,f_{js}))]$ ;

where the Kronecker-delta $\delta(\tau_s,f_{js})$ equals 1 iff $\tau_s=f_{js}$, and is 0 otherwise; it may be expressed as

(4)    $\delta(\tau_s,f_{js}) = \tau_s f_{js}(1+ \tau_s f_{js})/2 = f_{js}(\tau_s+ f_{js})/2$ .

In the following V(n,m) denotes the set of all CNF-formulae with exactly n variables and m clauses, which additionally does not contain equal clauses and must not contain the trivial clause, i.e. all $l_s = 0$. Also, any $F \in V(n,m)$ contains all n variables or its negations by definition. Randomness is introduced by a probability distribution $\rho_F$. Thus for any random variable X on V(n,m)

(5) $\quad E[X] = \sum_{F \in V(n,m)} \rho_F X(F) \quad \text{and} \quad \sum_{F \in V(n,m)} \rho_F = 1 \;.$

In most computer experiments on random SAT, however, not formulae F but clauses C are selected at random and independently. The corresponding formulae F consisting of m such independently selected clauses may therefore contain less than all n variables and do not fall into V(n,m) but are elements of a larger sample space $V_<(n,m)$ which contains all formulae with m clauses and <u>up to n variables.</u> In the large n limit, however, this difference has no effect on the calculation of transition lines for the models under consideration. We shall therefore understand the summation in (5) as a summation over m independently chosen clauses with the effect that the probability density $\rho$ factorizes with respect to clauses. $\rho_F$ can then be represented by some function $\rho$ of the clause variable matrix in the following form:

(6) $\quad \rho_F = \prod_{j=1}^{m} \rho(f_{j1}, \ldots, f_{jn}) \;.$

*Choice of random variables.*

For fixed n, choose as a random variable

(7) $\quad H(F) = \prod (1 - T_\tau(F))$

with the product extending over all labels $\tau = (\tau_1, \ldots, \tau_n) \,\varepsilon\, \{1,-1\}^n$. Since a single vanishing factor $1 - T_\tau(F)$ is sufficient to make H vanish we have H = 0 iff F is SAT, and H = 1 iff F is UNSAT. Thus with

(8) $\quad G = 1 - H$

one can calculate prob(SAT) directly

(9)   $\text{prob(SAT)} = \text{prob}(G > 0) = E[G]$ ,

in principle. In reality this venture is doomed because, as is clear from (7), it involves evaluating the expectation of all higher order correlations $T_\tau(F)\ T_{\tau'}(F) T_{\tau''}(F)\ldots$ We shall therefore adopt a method which makes use of inequalities for prob(SAT) and turns out to be sufficient for the determination of the transitional behaviour.

Quite generally one can state the following properties for any random variable X. If X fulfills

(10)   $X(F) \geq 0$   and   $X > 0$ iff F is SAT

then one has $\text{prob(SAT)} = \text{prob}(X > 0) \leq E[X]$ from Markovs inequality. Furthermore, if

(11)   $\delta X^2 = (E[X^2] - E[X]^2)/E[X]^2$

denotes the relative variance of X, then from the Cauchy-Schwartz-inequality one can deduce $1/(1+ \delta X^2) \leq \text{prob}(X>0)$ . Therefore one has quite generally the following lower and upper bounds on the probability that a randomly selected F is satisfiable:

(12)   $1/(1+ \delta X^2) \leq \text{prob(SAT)} \leq E[X]$

for any random variable X which satisfies (10).
These two inequalities are central in determining the transition line. We will evaluate the two quantities $\delta X^2$ and $E[X]$ in the following for X = N = number of solutions, i.e.

(13)   $N = \sum_\tau T_\tau(F)$ ,

which is the standard choice for a random variable to investigate the phase transition in random K-SAT. For the unrestricted random SAT-model we will find that both quantities $E[N]$ and $\delta N^2$ can be bounded from above by some quantity $2^{n\Phi}$ and its inverse, for large n, such that in essence

(14)   $1/(1+ 2^{-n\Phi}) \leq \text{prob(SAT)} \leq 2^{n\Phi}$ .

The quantity $\Phi$ depends on m, n and other parameters of the model in question. Equation (14) tells us that for $\Phi < 0$ prob(SAT) can be made arbitrarily small with increasing n, whereas for $\Phi > 0$ prob(SAT) comes arbitrarily close to 1. In other words, $\Phi = 0$ defines a surface in the parameter space which separates regions of low ($\Phi < 0$) and high satisfiability ($\Phi > 0$).

*(n,m,p)- Model.*

In the following we will consider distribution functions $\rho_F$ where clauses are generated randomly and each literal is chosen to be $a$, $\bar{a}$ or $0$ independently with probability p, p and 1-2p, respectively. In terms of (6):

(15) $\quad \rho(f_{j1}, ..., f_{jn}) = \prod_{s=1}^{n} \rho(f_{js}); \quad \rho(f_{js}=1) = p = \rho(f_{js}=-1); \quad \rho(f_{js}=0) = 1-2p$.

With (13) and (3) – (5) the mean number of solutions can be calculated readily

(16) $\quad E[N] = 2^n (1-(1-p)^n)^m$.

Note that the probability for a random clause to be solved by an arbitrary assignment, given by $1-(1-p)^n$, replaces $1-2^{-K}$ in the case of K-SAT.

From (12) and (14) the upper bound on prob(SAT) is given by $2^{n\Phi}$, with

(17) $\quad \Phi_{ub} = 1 + c \ln(1-(1-p)^n)/\ln(2) \approx 1 - c(1-p)^n/\ln(2)$

with density $c = m/n$, and the approximation improves with increasing n.

In the case of K-SAT this leads to $\phi_{ub} = 1 - c/c_{cr}(K)$ with a critical density $c_{cr}(K) = -\ln(2)/\ln(1-2^{-K})$. In particular K=3 yields the well known upper bound $c_{cr}(3) = \ln(2)/\ln(8/7) \approx 5.19089$. Also, for K>>1 one finds $c_{cr}(K) = 2^K \ln(2)$.

For the unrestricted random SAT model (17) and (14) imply a strongly n-dependent upper bound on the critical density given by

(18) $\quad c_{ub} = -\ln(2)/\ln(1-(1-p)^n) \approx \ln(2)/(1-p)^n$.

How far is this threshhold from the transition line, if it exists? To answer this question we investigate the l.h.s. of (12). For the calculation of $\delta N^2$ one needs to evaluate

$$E[N^2] = \sum_{F \in V(n,m)} \rho_F (\sum_{\tau} T_{\tau}(F))^2 = E[N] + 2\mu_2$$

with the two point correlation function

(19) $\mu_2 := \sum_{F \in V(n,m)} \rho_F \sum_{\langle\tau,\tau'\rangle} T_\tau(F) T_{\tau'}(F)$

($\langle\tau,\tau'\rangle$ indicates all mutually different pairs $\tau'$, $\tau$). Since clauses are chosen independently, see equ. (6), $\mu_2$ can be written as

$\sum_{\langle\tau,\tau'\rangle} [\sum_{f_1,...,f_n} \rho(f_1, f_2,..., f_n)(1-\prod_{s=1}^{n}(1-\delta(\tau_s,f_s)))(1-\prod_{s=1}^{n}(1-\delta(\tau'_s,f_s)))]^m$. Inserting the distribution

(15) yields

$\sum_{\langle\tau,\tau'\rangle} [1 - 2(1-p)^n + \prod_{s=1}^{n}(1-2p + p(\tau_s\tau'_s + 1)/2)]^m$ .

Since there are $2^{n-1}\binom{n}{\sigma}$ possibilities to choose mutually different pairs $\tau,\tau'$ with hamming distance $\sigma$, i.e. $\sigma$ mismatches ($\tau_s \neq \tau'_s$), we end up with

(20) $\mu_2 = \sum_{\sigma=1}^{n}\binom{n}{\sigma} 2^{n-1} [1 - 2(1-p)^n + (1-2p)^\sigma(1-p)^{n-\sigma}]^m$ .

The term in brackets is the probability that a randomly chosen clause is satisfied by both members of an arbitrary pair of assignments with $\sigma$ mismatches. And since the missing $\sigma=0$ term equals $E[N]/2$, we have for the (n,m,p)-model:

(21) $E\{N^2\} = \sum_{\sigma=0}^{n}\binom{n}{\sigma} 2^n [1 - 2(1-p)^n + (1-2p)^\sigma(1-p)^{n-\sigma}]^m$

And for $1+ \delta X^2$

(22) $1+ \delta X^2 = 2^{-n} \sum_{\sigma=0}^{n}\binom{n}{\sigma} [1 + ((1-2p)^\sigma(1-p)^{n-\sigma} - (1-p)^{2n})/(1 - (1-p)^n)^2]^m$

With $x = (1-p)^n$, $Y = x/(1-x)$ and $\beta = (1-2p)/(1-p)$ (22) can be written

(23) $\quad 1+ \delta X^2 = 2^{-n} \sum_{\sigma=0}^{n} \binom{n}{\sigma} [1 + Y^2(\beta^\sigma/x-1)]^m$

Since we are interested in a lower bound on the inverse of this quantity (see equ. (12)) we neglect the -1 and replace the power by an exponential. Finally:

(24) $\quad 1+ \delta X^2 \leq 2^{-n} \sum_{\sigma=0}^{n} \binom{n}{\sigma} \exp(z \beta^\sigma) \; ; \; z := mY^2/x \;.$

Expanding the exponential and performing the sum over $\sigma$ we rewrite this equation as

(25) $\quad 1+ \delta X^2 \leq 1 + \sum_{k=1}^{\infty} z^k/k! (1/2+\beta^k/2)^n$

Now choose an arbitrary cutoff $1 \ll \omega-1 \ll n$ and split the k-sum into a finite section running from 1 to $\omega-1$ and an infinite section. In the finite section all factors $(1/2+\beta^k/2)^n$ are bounded from above by $(1/2+\beta/2)^n$, which is smaller than 1 for $p < \frac{1}{2}$ and therefore tends to zero exponentially in n; the factor is a polynomial of order $\omega$ in z. Now, if $z = O(n)$ (which we asume and will be shown to be consistent in the following) the whole term vanishes in the large n limit. In the infinite section all terms in the sum are dominated by $(1/2+\beta^\omega/2)^n$. Therefore, completing the infinite sum by the finite terms from $k=1$ to $k=\omega$:

(26) $\quad 1+ \delta X^2 < 1 + (1/2+\beta^\omega/2)^n e^z \approx 1 + \exp\{z - n(\ln(2) - \ln(1+\beta^\omega))\} \approx 1 + \exp\{z - n(\ln(2)\} \;.$

Thus from (12) and (17):

(27) $\quad (1+ 2^{-n\Phi})^{-1} < 1/(1+ \delta X^2) \leq \text{prob(SAT)}$

with $\phi = 1 - c/c_{cr}$, and $c_{cr} = \ln(2)(1-x)^2/x \approx \ln(2)/x = \ln(2)(1-p)^{-n}$.

We conclude that for large n prob(SAT) is arbitrarily close to 1 when $\phi = 1 - c/c_{cr} > 0$, i. e. $c < c_{cr}$. For $c > c_{cr}$ $\phi$ is negative, the l.h.s. of (27) is zero, but due to (12) and (16) prob(SAT) is exponentially small for m,n-values above the curve $c = c_{cr}$. Therefore it is exactly this line which separates the (m,n)-plane into regions of low and high probability, for large n.

*Extended (biased) model.*

One can generalize the foregoing results to a model with an asymmetry between variables a and their complements. All clauses are chosen from a sample space in which each variable $a_j$ occurs with probability p and its complement $\bar{a}_j$ with probability q. This corresponds to a probability density $\rho_F$, instead of (15)

(28)   $\rho(f_{js}= 1) = p \quad \rho(f_{js}= -1) = q \quad \rho(f_{js}= 0) = 1-p-q$

The case considered before corresponds to p=q.

Because of the asymmetry we can no longer assume that the number of clauses satisfying a given truth assignment $T_\tau$ is independent of $\tau$. It now depends on the number of -1 in the string representing $T_\tau$. The corresponding probability ( $\lambda$ = number of -1 in $T_\tau$ ) is

(29)   $P(n,\lambda) = 1 - (1-p)^\lambda (1-q)^{n-\lambda}$ .

Since there are $\binom{n}{\lambda}$ many $T_\tau$ with $\lambda$ many $-1$, we have instead of (16)

(30)   $E[N] = \sum_{\lambda=0}^{n} \binom{n}{\lambda} P(n,\lambda)^m = \sum_{\lambda=0}^{n} \binom{n}{\lambda} (1 - \alpha^\lambda (1-q)^n)^m$ ;  $\alpha \equiv (1-p)/(1-q)$

Note that p=q gives the result derived in the foregoing section.

Likewise one can evaluate the two point correlation function, see equation (19), for this non uniform and asymmetric model. To this end it is necessary to classify pairs of truth assignments with respect to their number of mismatches $\sigma$ <u>and</u> to whether their matching digits equal -1 ($\lambda$ many) or 1 (n−σ−λ many).

There are $2^{\sigma-1} \binom{n}{\sigma}\binom{n-\sigma}{\lambda}$ many different such pairs, and instead of (20) one gets

(31)   $\mu_2 = \sum_{\sigma=1}^{n} \sum_{\lambda=0}^{n-\sigma} 2^{\sigma-1} \binom{n}{\sigma}\binom{n-\sigma}{\lambda} P(n,\lambda,\sigma)^m$

where P is the probability that a clause satisfies both assignments; it is given by

(32) $P(n,\lambda,\sigma) = 1 - (1-p)^{\lambda}(1-q)^{n-\lambda-\sigma}\{(1-p)^{\sigma}+(1-q)^{\sigma}-(1-p-q)^{\sigma}\}$ .

$= 1 - \alpha^{\lambda}(1-q)^{n}\{\alpha^{\sigma} + 1 - (1-p-q)^{\sigma}/(1-q)^{\sigma}\}$

If we let $p > q$ without loss of generality then $\alpha < 1$ and each summand in (30) is bounded by $(1-(1-p)^n)^m$ from above, and consequently E[N] is bounded as before suggesting a UNSAT phase for

(33) $c > \ln(2)/(1-p)^n$.

However, with the general model (31), (32) it is difficult to derive a lower bound on prob(SAT) according to (14). Only in the section of the parameter space (p,q) given by

(34) $q \leq p \leq 1-(1-q)(1-y/2)^{1/n}$ ; $y \equiv (1-q)^n$

we can derive a meaningful bound corresponding to the uniform case p=q. (34) restricts p and q to a tiny strip just above the straight line p=q. As long as (34) holds we have
$1-y(\alpha^{\sigma+\lambda} + \alpha^{\lambda}) \leq 1-2y\alpha^n \leq (1-y)^2 \leq (2^{-2n}E[N]^2)^{1/m}$ and thus

(35) $1+ \delta X^2 \leq 2^{-n} \sum_{\sigma=0}^{n}\binom{n}{\sigma}2^{\sigma}\sum_{\lambda=0}^{n-\sigma}\binom{n-\sigma}{\lambda}[1+\frac{y\alpha^{\lambda}\beta^{\sigma}}{(1-y)^2}]^m$ ; $\beta \equiv (1-p-q)/(1-q)$

which can be brought into the form (24)

(36) $1+ \delta X^2 \leq 2^{-n} \sum_{\sigma=0}^{n}\binom{n}{\sigma}\exp(z\beta^{\sigma})$ ; $z \equiv my/(1-y)^2 \approx m(1-q)^n$

Arguments closely following the lines leading from (24) to (26), we can finally state
$1+ \delta X^2 \leq 1 + \exp(z - n\ln(2) + O(n\beta^{\omega}))$ with some arbitrary cutoff $\omega$. Finally we get for the asymmetric model

(37) $(1+ 2^{-n\,\Phi(q)})^{-1} \leq 1/(1+ \delta X^2) \leq \text{prob(SAT)} \leq 2^{n\,\Phi(p)}$

with $\phi(p)$ given by $1 - m/n(1-p)^n/\ln(2)$. Note that now we have different $\phi$ on each side of the inequality. If we define two critical densities

(38) $c_{lb} = \ln(2)/(1-q)^n$ and $c_{ub} = \ln(2)/(1-p)^n$

then (37) states that with high probability F is SAT for $m/n < c_{lb}$, and UNSAT for $m/n > c_{ub}$. While both bounds are exponentially large their distance is only of order $O(1)$, and their relative distance exponentially small.

$$(c_{ub} - c_{lb})/c_{ub} \approx \frac{1}{2}(1-q)^n \; ; \text{ due to condition (34).}$$

Equations (37) and (38) constitute one central result of this paper. In particular in the unbiased case $p = q$ they yield a sharp transition between SAT and UNSAT regions in the parameter space. That is because the two bounds coincide exactly in this case ; then $m = n\ln(2)(1-p)^{-n}$ defines planes in (n,m,p)-space which separate SAT and UNSAT regions for large n. One may also state the result as follows: since $2pn$ is the mean clause length $\kappa$ of the symmetric (unbiased) model we expect a phase transition to occur, when $c = m/n$ exceeds the critical density

$$(39) \quad c_{cr} = \ln(2)/(1-\kappa/2n)^n \approx \ln(2)\exp(\kappa/2).$$

(39) apparently supports earlier results on a similar model, which found that a random F is satisfied by a random assignment with high probability if $\ln(m) < \kappa/2$, and has a low probability to be solved by a random assignment when $\ln(m) > \kappa$, [3], as cited in [1]. The probability distribution in [3] differs somewhat from our distribution (15), however. This may be the reason, why the results are interpreted differently, as we will discuss in the last section. Finally we state, that the line $c_{cr}(\kappa)$ can be introduced easily into the diagram given in [1,9] which indicates polynomial solubility of random SAT for a wide range of algorithms in the $(c, \kappa)$-plane. Earlier results on (n,m,p)-models, [6,10], labeled "Goldberg", could not be placed in that diagram consistently.

*Summary and discussion.*

We have derived the transition line for the transition from satisfiability to unsatisfiability in the (m,n)-plane for random CNF formulae with m clauses, n variables and mixed clause length with mean value $\kappa=2np$. A lower and an upper bound on the critical ratio m/n were derived from inequalities involving the first and second moments of the random variable N = number of solutions. The method, when used for K-SAT, i.e. formulae with a restricted number K (or up to K) of literals per clause, fails because the second moment of N only gives an exponentially small lower bound on the solvability probability, as pointed out in [4]. The reason is that the probability for a random clause to be solved by two randomly chosen assignments with hamming distance $\sigma$ is given by $1- 2x + x((n-\sigma)/n)^K$ where $x = 2^{-K}$. It shows that for $\sigma = n/2$ solutions are uncorrelated. This is not the case in the (n,m,p)-model, where the corresponding function (see (21) ff) is given by $1 - 2x + x[(1-2p)/(1-p)]^{\sigma}$, with $x = (1-p)^n$. Here long range interactions persist in the presence of disorder, showing up in the markedly different $\sigma$-dependence of this probability. Stated more technically, the complication does not arise at least in the symmetric case p=q, because the unrestricted model may be considered as an ensemble of K-SAT models with all values of K "summed up" before the large n limit is performed. Therefore one gets a different behaviour of correlation functions, which shows up most clearly in the n-dependence of most quantities on $m(1-p)^n \approx m\exp(-\kappa/2)$. Note that this quantity is the mean number of clauses unsatisfied by a randomly drawn assignment. It differs markedly from the mean number of trivial clauses (all $l_j = 0$), given by $m(1-2p)^n$, such that in the neighborhood of the transition curve the density of trivial clauses, $((1-2p)/(1-p))^n$, is exponentially small. Thus the fact that a random F is solved with high probability by a random assignment for $\ln(m) < \kappa/2$, cannot be explained by a high number of trivial clauses, as stated in [3]. This fact is probably the reason, why (n,m,p)-models have been abandoned earlier on, before they were studied theoretically in depth. It becomes apparent also, when the critical curve $c_{cr} \approx \ln(2)\exp(\kappa/2)$ is inscribed in a $(\kappa,c)$-diagram showing the performance of different SAT algorithms on random SAT models like the one discussed here (see [1,9] for that diagram). Although the curve marks a big portion of the region enclosed by $m/n > 1$ and $\kappa < \sqrt{m}$ as UNSAT, many points in that region are solved in polynomial time by probe backtracking [11].

As an additional result, we would like to put forward a scaling function for the (n,m,p)-model. From computer experiments on random K-SAT models one can try to derive a universal function which allows to determine the values of prob(SAT) for different n once the c-axis is properly rescaled: prob(SAT;K) = $u(\gamma_K(n)(c-c_{crK}))$. (see e.g. [5], also for further references). There is no theoretical derivation of the scaling factors $\gamma_K(n)$ as far as we know. The considerations in the foregoing section show that for the (n, m, p)-model

(40) $\quad u(x) = (1+2^x)^{-1}$

with $x := - n\phi(p)$ (see (37) ff) is a good approximation to the fundamental function u(x) of this model, at least for large n. Writing $\phi=1- c/c_{cr}(n)$ shows that the transition width is O(1/n) when c is properly rescaled.

We also note that a good approximation to the transition curve can be derived from a "mean field"-approximation, where correlations between different solutions are neglected and their effect is approximated by a "mean field" $E[T_\tau]$. More specifically, equ. (9) for the solvability probability prob(SAT) is approximated by

$$\text{prob(SAT)} = E[1 - H] \approx 1 - \prod (1 - E[T_\tau]) =: \text{prob}_{mf}(\text{SAT}) = 1 - \{1 - (1-(1-p)^n)^m\}^t$$

for the unbiased (n,m,p)-model, with $t = 2^n$. It is not difficult to see that

with $\Psi = t(1-(1-p)^n)^m = \exp(n\ln(2)(1 - c/c_{mf}))$, and $c_{mf} = -\ln(2)/\ln(1-(1-p)^n)$ we have

$$\Psi/(1 + \Psi) \leq \text{prob}_{mf}(\text{SAT}) \leq \Psi.$$

For c smaller than this critical density $\Psi^{-1}$ is exponentially small (in n) and $\text{prob}_{mf}(\text{SAT})$ is exponentially close to 1, according to the left inequality. For c larger than $c_{mf}$ $\Psi$ and thus $\text{prob}_{mf}(\text{SAT})$ is exponentially small, according to the right inequality. Thus $c_{mf}$ is the critical density in mean field approximation. Since $(1-p)^n \ll 1$ it is close to the rigorous result, equ. (39), derived for the (n,m,p)-model. Note that the result can be derived from a scaling function $\text{prob}_{mf}(\text{SAT}) = (1 + \Psi^{-1})^{-1}$, as indicated in equ. (40).

Finally, we would like to point out that random K-SAT models suffer quite generally from excluding large portions of the solution space. This can be seen as follows. Any boolean formula F with n variables can have at most $t=2^n$ solutions $T_j$, because there are no more different assignments. There are $2^t$ many possible combinations of solutions, i.e. the power set of $\{T_1, \ldots, T_t\}$. A set of Boolean formulae, which for any member $\{T_j, T_k, \ldots, T_l\}$ of the power set contains at least one F which is satisfied by all (say $\lambda$) $T \varepsilon \{T_j, T_k, \ldots, T_l\}$, must contain at least

$$(41) \quad v_{min} = 1 + n + \ldots + \binom{n}{\lambda} + \ldots + 1 = 2^t = \exp(2^n \ln(2))$$

logically inequivalent formulae. However, the set of all K-SAT CNF-formulae with n variables and m clauses, contains less than $\sum_{m=0}^{\theta} \binom{\theta}{m}$ many formulae, where $\theta = \binom{n}{K}$, (allowing for $m > \binom{n}{K}$ solely generates trivial duplicates). Thus the maximum number of logically inequivalent K-SAT CNF-formulae is bounded by

$$(42) \quad v_{K\text{-SAT}} = 2^\theta = \exp(\binom{n}{K} \ln(2)) \leq \exp(\varphi_K(n))$$

where $\varphi_K$ is a polynomial of order K in n. Therefore $v_{K\text{-SAT}}$ is much smaller than $v_{min}$ for large n. Thus the chance that a random K-SAT CNF-formula is satisfied by a given set of assignments is overexponentially small. In this sense the (n,m,p)-model discussed in this paper provides a much richer sample space, since its cardinality is of order $\exp(3^n) > v_{min}$. We believe that this is the proper reason for the markedly different critical behaviour of the(n,m,p)-model.